# Aftershocks in Modern Perspectives: Complex Earthquake Network, Aging, and Non-Markovianity

Sumiyoshi ABE [1]* and Norikazu SUZUKI [2]

[1] Department of Physical Engineering, Mie University, Mie 514-8507, Japan
[2] College of Science and Technology, Nihon University, Chiba 274-8501, Japan

Abstract

The phenomenon of aftershocks is studied in view of science of complexity. In particular, three different concepts are examined: (i) the complex-network representation of seismicity, (ii) the event-event correlations, and (iii) the effects of long-range memory. Regarding (i), it is shown the clustering coefficient of the complex earthquake network exhibits a peculiar behavior at and after main shocks. Regarding (ii), it is found that aftershocks experience aging, and the associated scaling holds. And regarding (iii), the scaling relation to be satisfied by a class of singular Markovian processes is violated, implying the existence of the long-range memory in processes of aftershocks.

______________________________________

\* Corresponding author. E-mail address: suabe@sf6.so-net.ne.jp

# 1. INTRODUCTION

Seismicity has been attracting continuous interest of physicists from the viewpoint of complex systems science. A reason behind this stream may be due to the fact that it is characterized by two classical laws indicating how seismicity is a complex phenomenon: the Gutenberg-Richter law (Gutenberg and Richter 1949) and Omori law (Omori 1894, Utsu 1961). The former states the scaling relation between cumulative frequency of event occurrence and released energies, and the latter tells power-law decay of occurrence of aftershocks following a main shock.

Although seismology has a long tradition, microscopic dynamics governing seismicity is still largely unknown. In such a situation, it is important to investigate the properties of correlations. In recent years, some empirical laws concerning spatiotemporal correlations have been investigated. Both 3-dimensional distance (Abe and Suzuki 2003) and time interval (calm time or inter-occurrence time) (Corral 2004, Abe and Suzuki 2005a) between two successive earthquakes were found to obey definite statistical laws that significantly deviate from Poissonian. In other words, two successive events are correlated at least at the statistical level with high significance. In addition, it is known (Steeples and Steeples 1996) that an earthquake can trigger the next one that can be more than 1000 km away. Thus, the event-event correlation length may be divergently large, indicating a strong similarity to critical phenomena. Accordingly, we propose to frame the working hypothesis that two successive events are statistically correlated. (This hypothesis is not satisfied by pairs that are not causal in terms of the propagation speed of seismic waves, for example. However, such pairs are statistically not significant if a large number of events are considered.)

In this paper, we report recent discoveries about the physics of aftershocks. Specifically, they are relevant to (i) evolution of *complex earthquake networks*, (ii) *aging* and *scaling*, and (iii) long-range memory, i.e., the *non-Markovian* nature. All these issues are deeply concerned with correlations between aftershocks.

The present paper is organized as follows. In Section 2, first, the concept of earthquake network and the procedure of its construction are explained in detail. Then, it is shown how the clustering property of the earthquake network evolves in time before and after a main shock. In Section 3, the event-event correlations are considered. The two-point correlation function defined there is found to exhibit a peculiar non-stationary behavior, termed *aging*. A glassy dynamics aspect of seismicity of aftershocks is discussed. In Section 4, the non-Markovian nature of aftershocks is studied. It is shown that the scaling relation to be satisfied by a class of singular Markovian stochastic processes is violated by aftershocks. Section 5 is devoted to concluding remarks. The data analyzed in this paper is the one taken from California (http://www.data.sces.org/). However, we emphasize that the results presented here were reconfirmed by other datasets such as the Japanese one.

## 2.   COMPLEX EARTHQUAKE NETWORK AND ITS EVOLUTION

In contemporary statistical mechanics, the concept of complex networks appears as a powerful tool for quantifying the degree of complexity of a system/phenomenon. A network tells us about basic architecture underlying a complex system. In a recent work (Abe and Suzuki 2004a), this concept has been introduced into seismology. In Subsection 2.1, we explain the procedure of constructing an earthquake network in detail. In Subsection 2.2, we succinctly summarize the known fundamental properties of an earthquake network as a complex network. Then, in Subsection 2.3, which is the main part of this section, we discuss how time evolution of the clustering coefficient of an earthquake network can characterize aftershocks as well as a main shock.

**2.1 Earthquake network and its complexity**

Here, we explain the procedure of constructing an earthquake network originally presented in (Abe and Suzuki 2004a). First, we divide a geographical region under consideration into cubic cells. We regard a cell as a vertex of a network if earthquakes with any values of magnitude (above a certain detection threshold) occurred therein. Then, we link two vertices of successive events by an edge. If two events successively occur in the same cell, then we attach a tadpole (i.e., a self-loop) to that vertex. These edges and tadpoles represent the event-event correlations in accordance with the working hypothesis mentioned in the preceding section. A useful method we employ here for practically setting up the cells and identifying a cell for each earthquake is as follows. Let $\theta_0$ and $\theta_{max}$ be the minimal and maximal values of latitude of the whole region, respectively. Similarly, let $\phi_0$ and $\phi_{max}$ be the minimal and maximal values of longitude. Define $\theta_{av}$ as the sum of the values of latitude of all the events divided by the number of events. The hypocenter of the $i$th event is denoted by $(\theta_i, \phi_i, z_i)$, where $\theta_i$, $\phi_i$, and $z_i$ are the values of latitude, longitude and depth, respectively. The north-south distance between $(\theta_0, \phi_0)$ and $(\theta_i, \phi_i)$ reads $d_i^{NS} = R \cdot (\theta_i - \theta_0)$, where $R (\cong 6370 \, \text{km})$ is the radius of the Earth. On the other hand, the east-west distance is given by $d_i^{EW} = R \cdot (\phi_i - \phi_0) \cdot \cos \theta_{av}$. (In these expressions, all the angles should be described in the unit of radian.) The depth is simply $d_i^D = z_i$. Starting from the point $(\theta_0, \phi_0, z_0 \equiv 0)$, divide the region into cubic cells with a given value of the cell size $L[\text{km}] \times L[\text{km}] \times L[\text{km}]$. Then, the cell of the $i$th event can be identified by making use of $d_i^{NS}$, $d_i^{EW}$, and $d_i^D$.

The above procedure allows us to map, in an unambiguous way, a given seismic time series to a growing stochastic network, which is an earthquake network that we have been referring to.

Several comments on the above-mentioned construction procedure are in order. First of all, it contains a single parameter: the cell size, which can be seen the scale of coarse graining. All earthquakes occurred in a given cell are identified. It is important to clarify how the properties the network depend on it. This point has thoroughly been discussed in a recent work (Abe and Suzuki 2009c). Secondly, an earthquake network is a directed network. However, directedness is irrelevant to statistical analysis of connectivity (or, degree, i.e., the number of edges attached to the vertex under consideration) needed for examining the scale-free property, since by construction the in-degree and out-degree are identical for each vertex except the initial and final vertices in analysis. So, they need not be distinguished. That is, vertices except the initial and final ones have the even-number values of connectivity. However, directedness is important, when the period distribution (i.e., the waiting event-time distribution) is considered (Abe and Suzuki 2005b). Thirdly, a full directed earthquake network should be reduced to a simple undirected network, when its small-world property is examined. There, tadpoles are removed, and each multiple edge is replaced by a single edge (see Figure 1). [The standard small-world network (Watts and Strogatz 1998) is simple and undirected.] We note that, according to our examinations, gross properties of a network do not change, although numerical values of characteristics of a network generically depend slightly on how cells are set up. Lastly, we mention that the cell size $L$ is supposed to be typically a few km or larger, in view of the smallest fault size as well as emergence of universalities of the network characteristics (Abe and Suzuki 2009c).

**2.2 Earthquake network as complex network**

In a series of our studies, we have discovered by analyzing real seismic data that an earthquake network is a complex network.

In the work (Abe and Suzuki 2004a), we have found that a full earthquake network is scale-free. That is, the probability $P(k)$ of finding vertices with connectivity, $k$, obeys a power law (Barabási and Albert 1999): $P(k) \sim k^{-\gamma}$, where $\gamma$ is a positive constant. A remarkable empirical fact is that aftershocks following a main shock tend to return to the cell of the main shock, geographically. This makes the vertex of the main shock a hub of the earthquake network. And, consequently, the network becomes heterogeneous.

In the works (Abe and Suzuki 2004c, Abe and Suzuki 2006a), we have shown that a simple network reduced from a full earthquake network is of the small-world type. Two main features of a small-world network are as follows (Watts and Strogatz 1998). One is that the average path length (i.e., the number of edges between two chosen vertices) is small. The other is that the clustering coefficient is much larger than that of the Erdös-Rényi random graph (Bollobás 2001). Here, the clustering coefficient $C$ of a simple network with $N$ vertices is defined as follows (Watts and Strogatz 1998):

$$C = \frac{1}{N}\sum_{i=1}^{N} c_i . \qquad (1)$$

$c_i$ appearing on the right-hand side is given by $c_i \equiv$ (number of edges of the $i$th vertex and its neighbors)$/[k_i(k_i-1)/2]]$ with $k_i$ being the connectivity of the $i$th vertex. It is calculated also as follows. Let $A$ be the symmetric adjacency matrix of a simple network. Its element $(A)_{ij}$ is 1 (0), if the $i$th and $j$th vertices are linked (unlinked). The diagonal elements of $A$ are zero, since a simple network has no tadpoles. Then, $c_i$ is also written as follows:

$$c_i = \frac{e_i}{e_i^{\max}} , \qquad (2)$$

where $e_i = (A^3)_{ii}$, and $e_i^{\max} = k_i(k_i-1)/2$ with $k_i = \sum_{j=1}^{N}(A)_{ij}$, which is nothing but the maximum value of $e_i$. $c_i$ quantifies the tendency that two neighboring vertices of the $i$th vertex are linked together (i.e., forming a triangle). By definition, $C$ takes a value between 0 and 1. As pointed out in the work (Watts and Strogatz 1998), $C$ of a small-world network is much larger than that of the corresponding Erdös-Rényi random graph: $C >> C_{random}$, where $C_{random} = <k>/N << 1$ with $<k>$ being the average value of connectivity of the random graph.

In fact, a reduced simple earthquake network has a small value of the average path length. For the number of vertices about 27000, it is less than 4 (Abe and Suzuki 2004c, Abe and Suzuki 2006a). Also, *C* of an earthquake network is $10^3 \sim 10^4$ times larger than $C_{random}$. Therefore, an earthquake network is a small-world network.

In addition to scale-freeness and small-worldness, earthquake network has further remarkable properties. Among others, the hierarchical organization (Ravasz and Barabási 2003) and mixing property (Newman 2002) should be emphasized. In the work (Abe and Suzuki 2006b), it is found that an earthquake network is, in fact, hierarchically organized and possesses assortative mixing (implying that a hub tends to be linked to other hubs rather than vertices with small values of connectivity). Also, spectral analysis shows (Abe and Suzuki 2009b) that an earthquake network is locally tree-like, that is, $c_i$ in eq. (2) is small if the *i*th vertex is a hub. Furthermore, there exists finite data-size scaling for the clustering coefficient (Abe, Pastén, and Suzuki 2011).

Closing this subsection, we stress the following point. Any seismic data may be incomplete due to errors, detection threshold, and so on. However, the basic properties of earthquake network are not affected by such incompleteness. As known in the literature (Albert *et al*. 2000), complex networks are highly robust to random failure.

## 2.3 Evolution of clustering coefficient and aftershocks

Now, let us see, as an example of applications of an earthquake network, how aftershocks as well as a main shock can be characterized in a peculiar way (Abe and Suzuki 2007). In particular, here we discuss time evolution of the clustering coefficient $C$ in eq. (1) around the occurrence times of some well-known main shocks.

We construct the earthquake network from every 240-hours (10-days) interval. Here, we specifically focus our attention on two celebrated main shocks: the Joshua Tree Earthquake (M6.1 on April 23, 1992) and the Hector Mine Earthquake (M7.1 on October 16, 1999). Here, the cell size is taken to be 5 km $\times$ 5 km $\times$ 5 km. The result is presented in Figure 2. There, we see a remarkable common behavior. The clustering coefficient stays stationary before the main shock, suddenly jumps up at the main shock, and then slowly decays to return to a stationary value again with some fluctuations. In the work (Abe and Suzuki 2007), it is shown that the decay process, which takes place during the interval of aftershocks, obeys a power law. Thus, the clustering coefficient characterizes aftershocks as well as a main shock in a novel way.

The above discussion is nothing but an example of possible applications. The complex network approach to seismicity is still at an infant stage, and a lot of issues are yet to be investigated.

## 3. AGING AND SCALING: ARE AFTERSHOCKS GOVERNED BY GLASSY DYNAMICS?

Regarding aftershocks, the following two points should be noted. A) The stress distribution at faults has a complex landscape. A main shock instantaneously releases a huge amount of energy with quenching the disorder of the stress distribution, in analogy with super-cooling. Then, it reorganizes the stress distribution. The "system" changes its energy state from one local minimum to another. Such a transition may be regarded as an aftershock. B) The relaxation of the system to a stationary (or, quasi-equilibrium) state is very slow according to the power-law nature of the Omori law (Omori 1894, Utsu 1961).

Keeping A) and B) in mind, here we discuss the concept of event-event correlations of aftershocks. Let $\{t_0, t_1, t_2, ..., t_{M-1}\}$ be a sequence of the occurrence times of aftershocks, where $t_0$ is the occurrence time of a chosen initial event of the sequence of aftershocks. Such a sequence defines a point process, and the *n*th occurrence time, $t_n$, is a random variable labeled by the number $n \, (= 0, 1, 2, ..., M-1)$, which is referred to as *event time*. The fundamental quantity is the following *event-event correlation function* proposed in the work (Abe and Suzuki 2004b):

$$C(m,n) = \frac{<t_m t_n> - <t_m><t_n>}{\sqrt{\sigma_m^2 \sigma_n^2}}, \tag{3}$$

where $<\bullet>$ is given by the *event-time average*: $<t_m> = (1/M) \sum_{k=0}^{M-1} t_{m+k}$, $<t_m t_n> = (1/M) \sum_{k=0}^{M-1} t_{m+k} \, t_{n+k}$, and $\sigma_m^2 = <t_m^2> - <t_m>^2$. If the process (i.e., the sequence) is non-stationary, this quantity depends on two event times, *m* and *n*, in general. It is convenient to introduce the *waiting event time*, $n_W$, to rewrite eq. (3) as

$$C(n + n_W, n_W). \tag{4}$$

If the process is stationary, the quantity in eq. (4) does not depend on $n_W$. Clearly, $C(n+n_W, n_W) = 1$ at $n=0$. This quantity might have a complicated $n_W$-dependence, in general. However, it actually turns out to exhibit a very special $n_W$-dependence.

We have analyzed the event-event correlation function of the aftershocks following the Landers Earthquake (M7.3 on June 28, 1992) as a typical example. In Figure 3, we present the plots of $C(n+n_W, n_W)$ for several values of $n_W$. A remarkable feature is observed there. The larger $n_W$ is, the slower $C(n+n_W, n_W)$ decreases. No crossing of the curves occurs. This phenomenon is called aging in statistical mechanics. It implies that the system has its own clock, recording its "intrinsic internal time".

Another important point is the existence of scaling. It is possible to collapse all curves in Figure 3 to a single curve (see Figure 4):

$$C(n+n_W, n_W) = \tilde{C}(n/f(n_W)), \qquad (5)$$

where $\tilde{C}$ is a scaling function. $f(n_W)$ is found to have the form

$$f(n_W) = a(n_W)^\gamma + 1. \qquad (6)$$

$a = 1.37 \times 10^{-6}$ and $\gamma = 1.62$ for the scaling function in Figure 4.

In the work (Abe and Suzuki 2004b), the aftershocks associated with other main shocks are also analyzed, and the same result as the above one is obtained. In addition, outside of the intervals of aftershocks, the aging phenomenon is not observed. Thus, aging characterizes aftershocks.

Now, combining the aging and scaling as well as the points A) and B) mentioned in the beginning of this section, we conclude that the dynamics governing aftershocks is highly similar to *glassy dynamics* (Fischer and Hertz 1991).

4. **NON-MARKOVIAN NATURE**

In a Markovian process, transition of a system from one state to another can basically be understood in terms of local fluctuations. In a non-Markovian process, on the other hand, such a local picture cannot apply because of the presence of the long-range memory, i.e., temporal non-separability of events. Non-Markovianity signals complexity of a system.

Given finite seismic data, it is generically difficult to determine if such a stochastic process is Markovian or not. However, there is a rigorous mathematical method of determining it for a class of singular point processes, which has been discussed in the problems of laser cooling of atoms (Bardou *et al*. 2002). We apply such a method to examining how the process of aftershocks is non-Markovian (Abe and Suzuki 2009a).

What is of crucial importance is the existence of the scaling relation in a class of singular Markovian processes. It is related to two basic quantities in the processes. One is the time-interval distribution, $P(t)$, of two successive aftershocks, and the other is the rate of event occurrence (i.e., temporal mean density of aftershocks), $S(t) = [N(t+\Delta t) - N(t)]/\Delta t$, where $N(t)$ stands for the number of events occurred in the time interval $[0, t]$. If a process is Markovian, then holds the following equation (Bardou *et al*. 2002):

$$S(t) = P(t) + \int_0^t dt'\, P(t-t')\, S(t'), \tag{7}$$

which can be derived from the Kolmogorov forward equation (Barndorff-Nielsen *et al.* 2000). Since the second term on the right-hand side is a convolution integral, it is convenient to perform the Laplace transformations of the both sides. Then, we obtain

$$L[S](s) = \frac{L[P](s)}{1 - L[P](s)}, \tag{8}$$

where $L[f](s) \equiv \int_0^\infty dt\, e^{-st} f(t)$. Consider a singular processes, in which both $P$ and $S$ decay as a power law

$$P(t) \sim \frac{1}{t^{1+\mu}}, \tag{9}$$

$$S(t) \sim \frac{1}{t^p}, \tag{10}$$

for a large value of $t$. If the exponents, $\mu$ and $p$, satisfy

$$0 < \mu < 1, \tag{11}$$

$$0 < p < 1, \tag{12}$$

then

$$L[P](s) \sim 1 - \alpha s^\mu, \tag{13}$$

$$L[S](s) \sim s^{p-1}, \tag{14}$$

for a small value of $s$, where $\alpha$ is a positive constant. Therefore, from eq. (8) it follows that

$$p + \mu = 1. \tag{15}$$

This is the scaling relation to be satisfied by singular Markovian processes with $P$ and $S$ obeying eqs. (9)-(12).

Note that eq. (10) precisely describes the Omori law for aftershocks. Therefore, a question arising is if the time-interval distribution, $P$, also obeys a power law for aftershocks. The answer to this question turns out to be affirmative, as we shall see below.

As in the preceding section, we analyze the aftershocks of the Landers Earthquake. We set the spatial window with size, 100 km (east-west) and 100 km (north-south) centered at the epicenter of the Landers Earthquake ($34°12.00$ N latitude, $116°26.22$ W longitude), and 100 km in depth. We consider 600 days after the main shock, during which there occurred 34783 events in this windowed region.

Figure 5 is the plot of $S(t)$. There, one sees that the Omori law holds well. Figure 6 is the plot of $P(t)$. One certainly recognizes the behavior in eq. (9). However, since the scaling regime is not so large, a careful examination is needed for determining the value of the exponent, $\mu$. To do so, we employ the method of maximum likelihood estimation (Goldstein *et al*. 2004, Newman 2005).

Table 1 summarizes the result. As can be seen there, the Markovian scaling relation in eq. (15) is violated about 50%. Therefore, we conclude that the process of aftershocks following the Landers Earthquake is non-Markovian, possessing the long-range memory.

Finally, we mention that a similar result is obtained for aftershocks of other main shocks such as the Hector Mine Earthquake (Abe and Suzuki 2009a). We confidently believe that the process of aftershocks is non-Markovian, in general.

## 5.  CONCLUDING REMARKS

To summarize, we have surveyed the recent discoveries about the nature of aftershocks in view of science of complexity. We have seen how the complex network representation can reveal the novel features of seismicity. Aging and scaling of aftershocks are certainly remarkable. It is of extreme importance to further clarify how the unknown dynamics governing aftershocks is similar to (or, different from) glassy dynamics. We have also seen that processes of aftershocks are non-Markovian. All these results highlight the aspects of aftershocks as a complex phenomenon. Future investigations along these lines based on modern science will give clues to deeper understandings of the physics of seismicity.

Acknowledgment. S. A. was supported in part by a Grant-in-Aid for Scientific Research from the Japan Society for the Promotion of Science.

# References

Abe, S. and N. Suzuki (2003), Law for the distance between successive earthquakes, *J. Geophys. Res.* **108** (B2), 2113, DOI: 10.1029/2002JB002220.

Abe, S. and N. Suzuki (2004a), Scale-free network of earthquakes, *Europhys. Lett.* **65**, 581-586.

Abe, S. and N. Suzuki (2004b), Aging and scaling of earthquake aftershocks, *Physica A* **332**, 533-538.

Abe, S. and N. Suzuki (2004c), Small-world structure of earthquake network, *Physica A* **337**, 357-362.

Abe, S. and N. Suzuki (2005a), Scale-free statistics of time interval between successive earthquakes, *Physica A* **350**, 588-596.

Abe, S. and N. Suzuki (2005b), Scale-invariant statistics of period in directed earthquake network, *Eur. Phys. J. B* **44**, 115-117.

Abe, S. and N. Suzuki (2006a), Complex-network description of seismicity, *Nonlin. Processes Geophys.* **13**, 145-150.

Abe, S. and N. Suzuki (2006b), Complex earthquake networks: Hierarchical organization and assortative mixing, *Phys. Rev. E* **74**, 026113.

Abe, S. and N. Suzuki (2007), Dynamical evolution of clustering in complex network of earthquakes, *Eur. Phys. J. B* **59**, 93-97.

Abe, S. and N. Suzuki (2009a), Violation of the scaling relation and non-Markovian nature of earthquake aftershocks, *Physica A* **388**, 1917-1920.

Abe, S. and N. Suzuki (2009b), Scaling relation for earthquake networks, *Physica A* **388**, 2511-2514.

Abe, S. and N. Suzuki (2009c), Determination of the scale of coarse graining in earthquake networks, *Europhys. Lett.* **87**, 48008.

Abe, S., D. Pastén, and N. Suzuki (2011), Finite data-size scaling of clustering in earthquake networks, *Physica A* **390**, 1343-1349.

Albert, R., H. Jeong, and A.-L. Barabási (2000), Error and attack tolerance of complex networks, *Nature* **406**, 378-382.

Barabási, A.-L. and R. Albert (1999), Emergence of scaling in random networks, *Science* **286**, 509-512.

Bardou, F., J.P. Bouchaud, A. Aspect, and C. Cohen-Tannoudji (2002), Lévy Statistics and Laser Cooling, Cambridge University Press, Cambridge.

Barndorff-Nielsen, O.E., F.E. Benth, and J.L. Jensen (2000), Markov jump processes with a singularity, *Adv. Appl. Prob.* **32**, 779-799.

Bollobás, B. (2001), Random Graphs, Second Edition, Cambridge University Press, Cambridge.

Corral, Á. (2004), Long-term clustering, scaling, and universality in the temporal occurrence of earthquakes, *Phys. Rev. Lett.* **92**, 108501.

Fischer, K.H. and J.A. Hertz (1991), Spin Glasses, Cambridge University Press, Cambridge.

Goldstein, M.L., S.A. Morris, and G.G. Yen (2004), Problems with fitting to the power-law distribution, *Eur. Phys. J. B* **41**, 255-258.

Gutenberg, B. and C.F. Richter (1949), Seismicity of the Earth and Associated Phenomena, Princeton University Press, Princeton.

Newman, M.E.J. (2002), Assortative mixing in networks *Phys. Rev. Lett.* **89**, 208701.

Newman, M.E.J. (2005), Power laws, Pareto distributions and Zipf's law, *Contemporary Physics* **46**, 323-351.

Omori, F. (1894), On the after-shocks of earthquakes, *J. Coll. Sci. Imp. Univ. Tokyo* **7**, 111-200.

Ravasz, E. and A.-L. Barabási (2003), Hierarchical organization in complex networks, *Phys. Rev. E* **67**, 026112.

Steeples, D.W. and D.D. Steeples (1996), Far-field aftershocks of the 1906 Earthquake, *Bull. Seismol. Soc. Am.* **86**, 921-924.

Utsu, T. (1961), A statistical study on the occurrence of aftershocks, *Geophys. Mag.* **30**, 521-605.

Watts, D.J. and S.H. Strogatz (1998), Collective dynamics of 'small-world' networks, *Nature* **393**, 440-442.

# Figure and Table Captions

**Figure 1** Schematic descriptions of earthquake network: (a) a full directed network and (b) its reduced undirected simple network in the small-world picture. The vertices with the dotted lines indicate the initial and final events in analysis.

**Figure 2** Evolution of the (dimensionless) clustering coefficient for each 240 hours. The cell size is taken here as 5 km $\times$ 5 km $\times$ 5 km. The origins are adjusted to the moments of the main shocks: (a) the Joshua Tree Earthquake and (b) the Hector Mine Earthquake. The length of the total interval in (a) is shorter than that in (b). This is because another main shock (i.e., the Landers Earthquake) occurred on June 28, 1992 and makes it impossible to take the aftershock sequence following the Joshua Tree Earthquake so long.

**Figure 3** The plots of the event-event correlation function with respect to the event time, $n$, for several values of waiting event time, $n_W$. $n$ runs within the interval of the aftershocks following the Landers Earthquake. (Color online) $n_W = 0$ (black), 600 (red), 1200 (green), and 1800 (blue). All quantities are dimensionless.

**Figure 4** The data collapse of Figure 3 by the rescaling of the event time. All quantities are dimensionless.

**Figure 5** The log-log plot of $S(t)$ with the inverse dimension of time. The bin size for constructing the histogram is taken to be 2 days.

**Figure 6** The log-log plot of $P(t)$ with the inverse dimension of time. The bin size for constructing the histogram is taken to be 30 s.

**Table 1** The values of $p$ and $\mu$ for some different bin size for constructing the histograms in the case of the aftershocks following the Landers Earthquake. The errors, which are automatically calculated by the method of maximum likelihood estimation, are very small due to the large sample size.

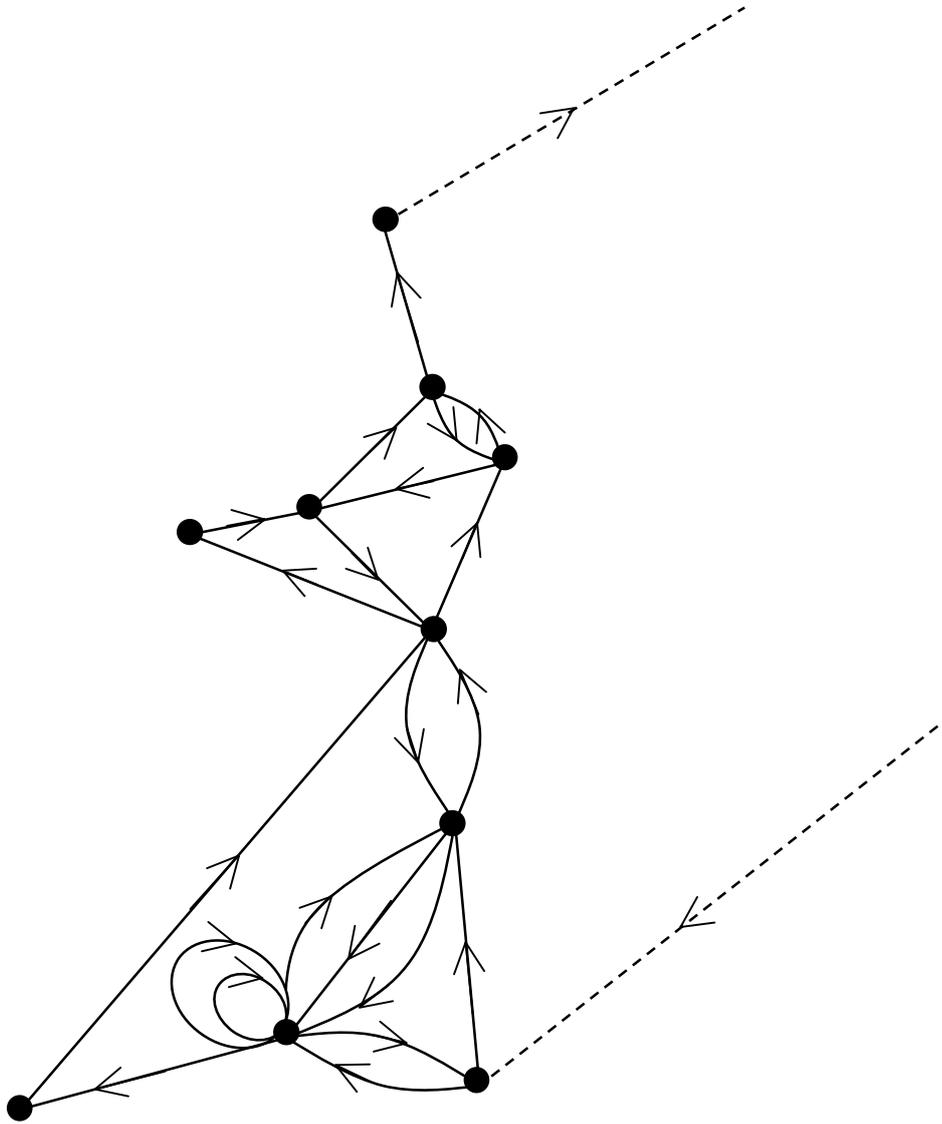

Figure 1 (a)

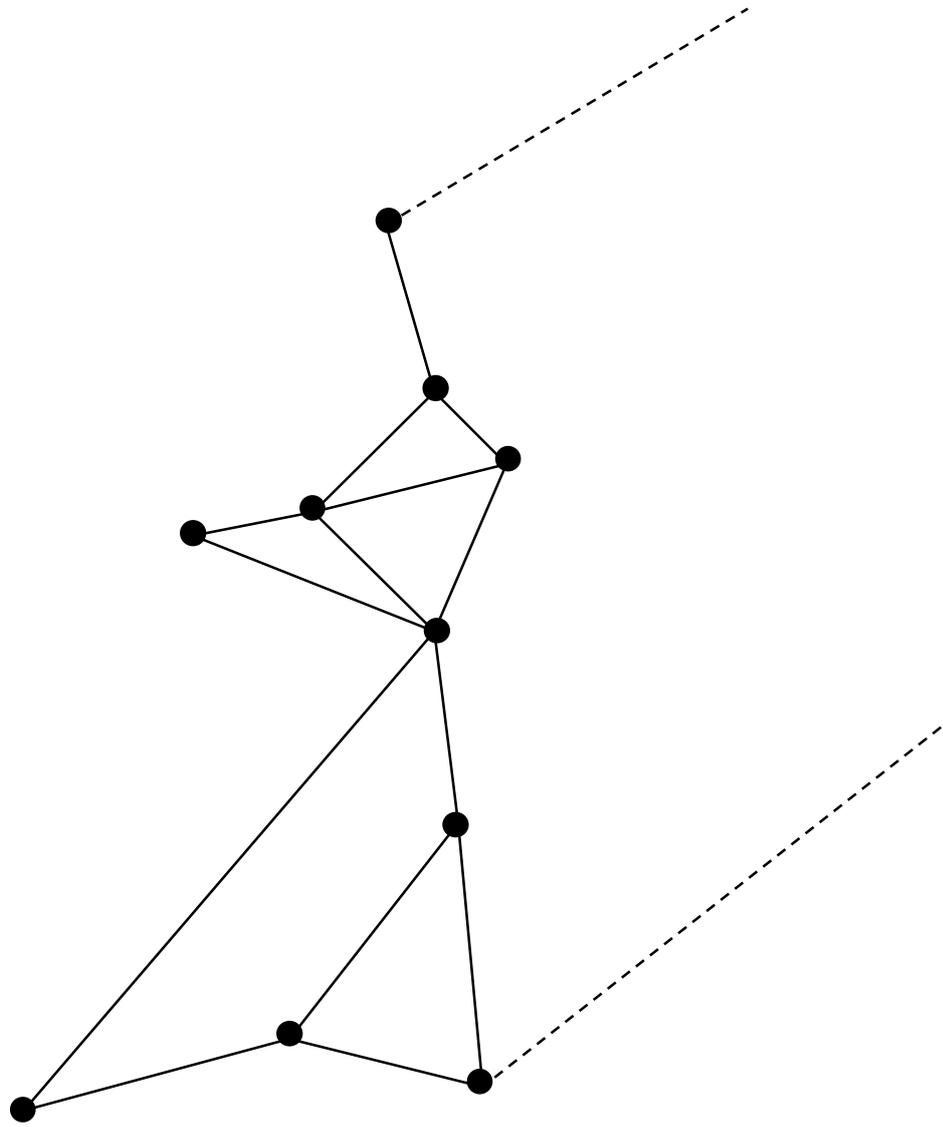

Figure 1 (b)

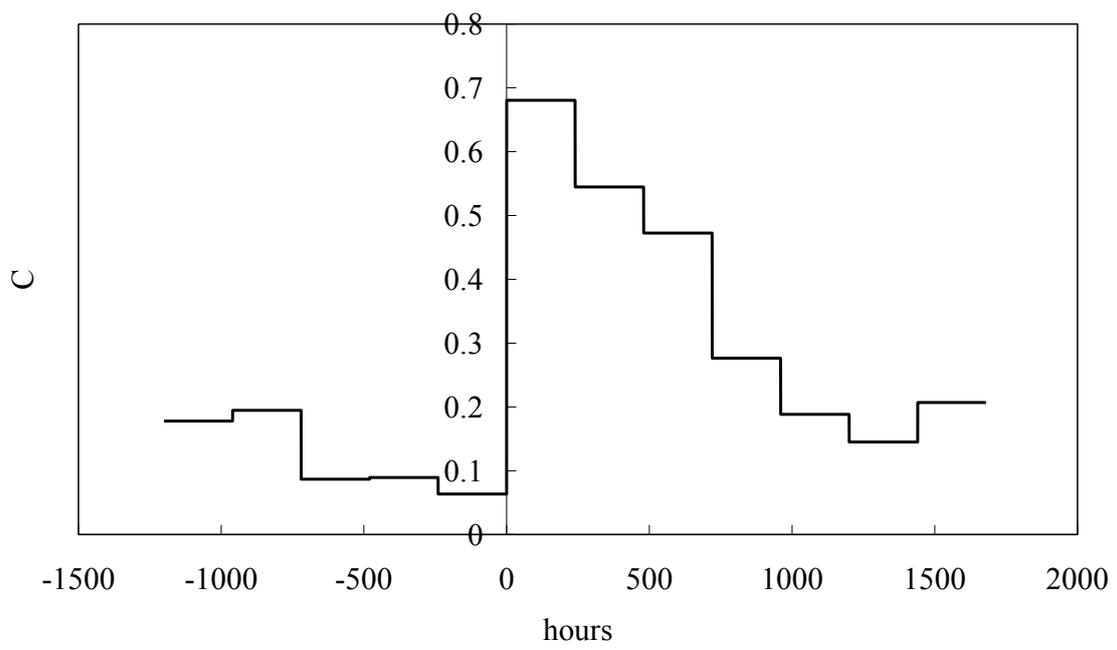

Figure 2 (a)

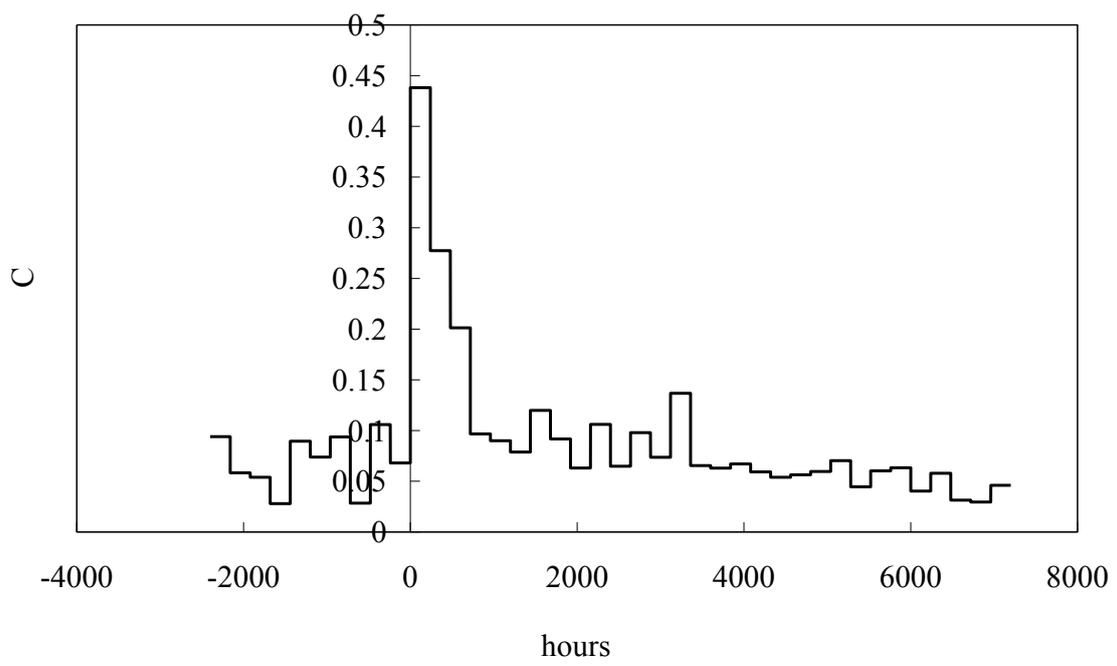

Figure 2 (b)

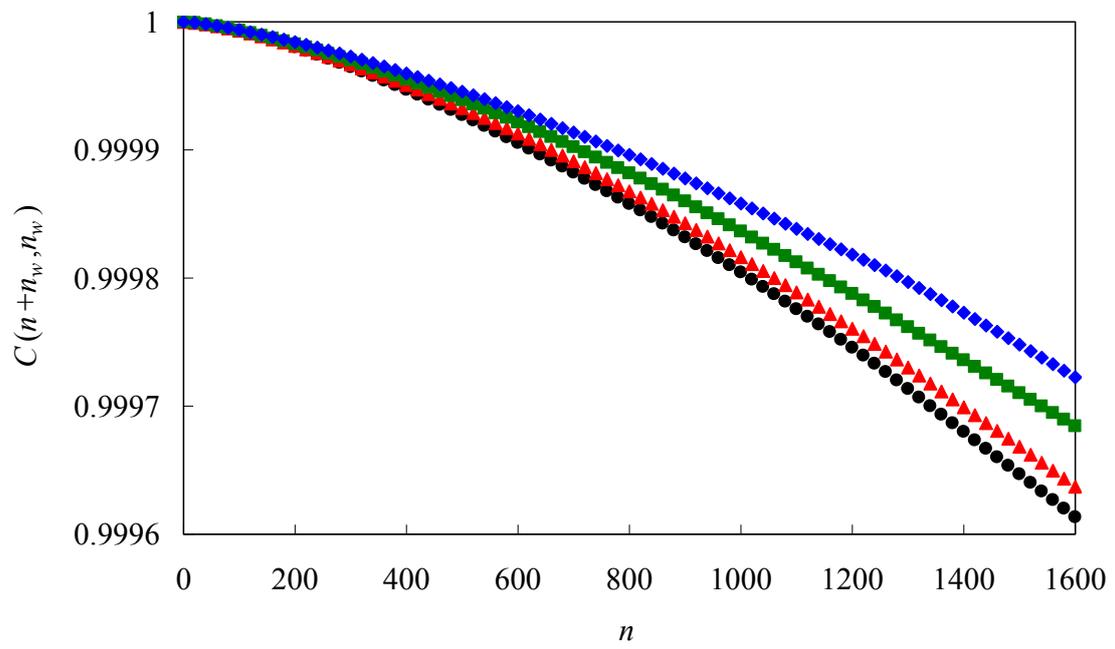

Figure 3

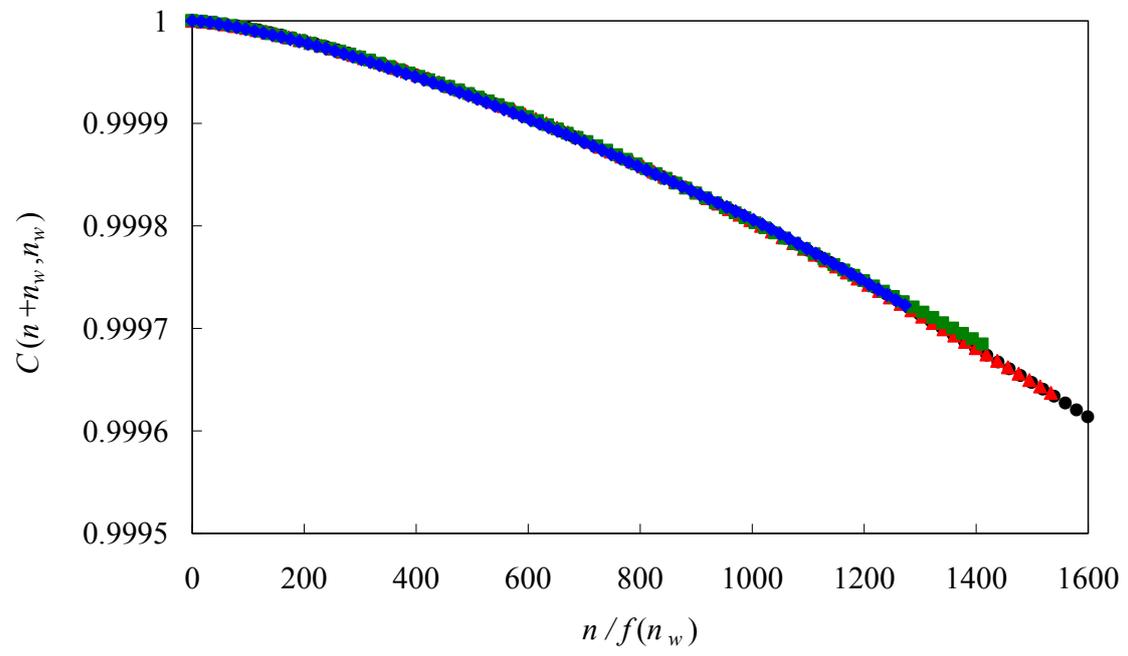

Figure 4

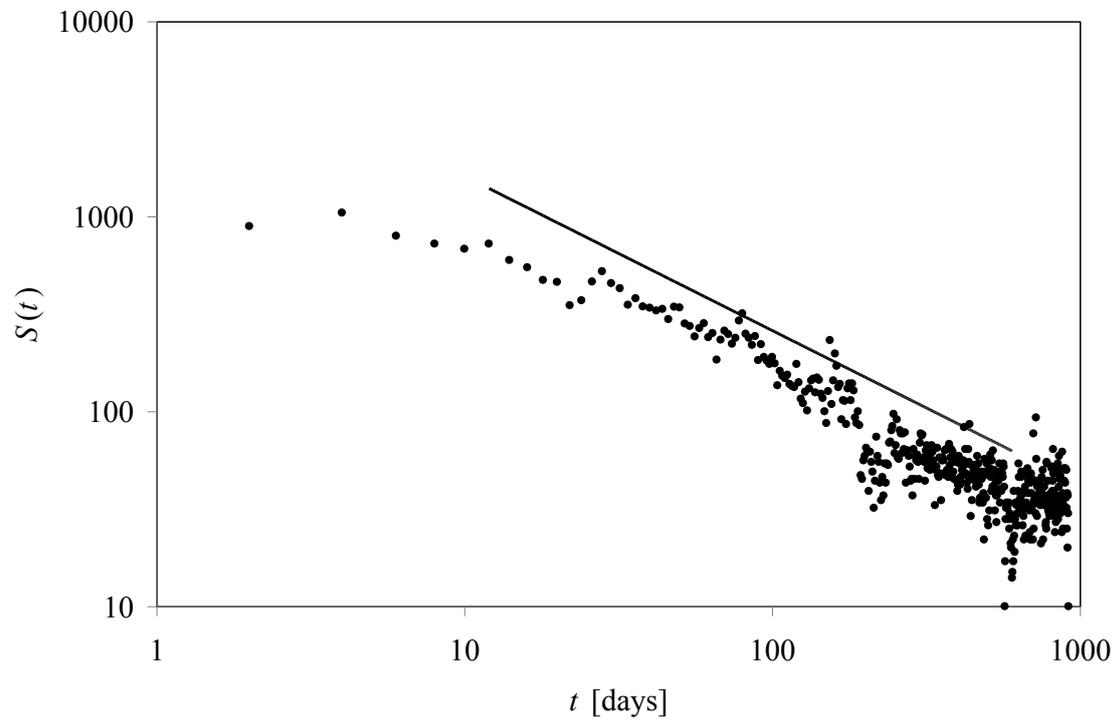

Figure 5

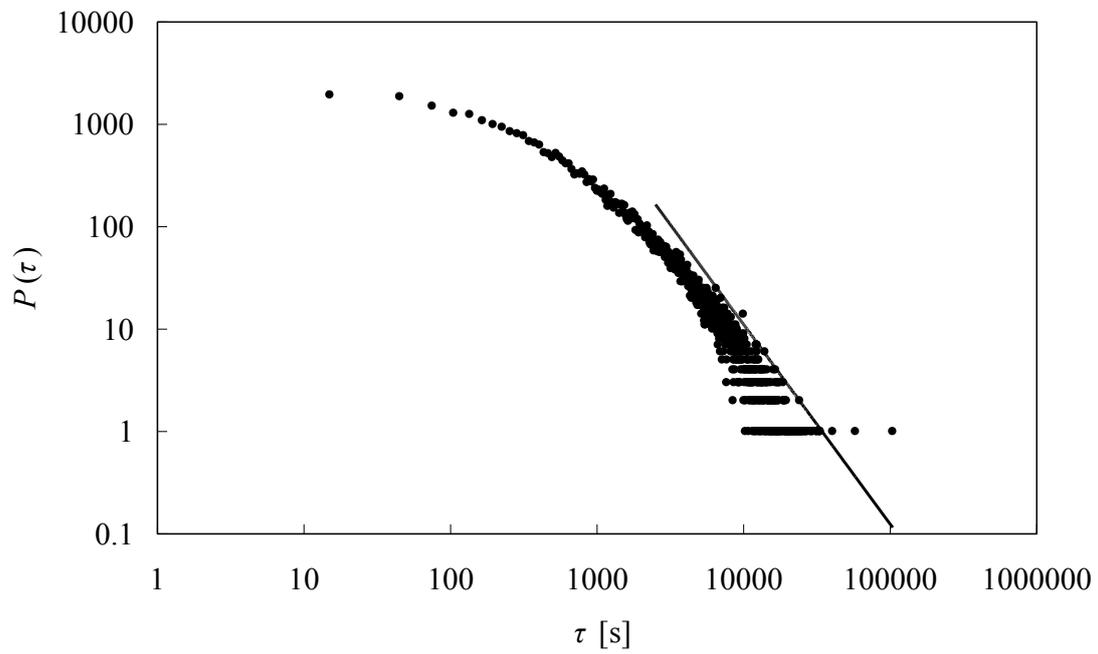

Figure 6

| bin size [days] | $p$ | bin size [s] | $\mu$ | $p + \mu$ |
|---|---|---|---|---|
| 0.5 | $0.788 \pm 0.004$ | 10 | $0.629 \pm 0.011$ | $1.417 \pm 0.015$ |
|  |  | 20 | $0.873 \pm 0.018$ | $1.661 \pm 0.021$ |
|  |  | 30 | $0.950 \pm 0.017$ | $1.738 \pm 0.021$ |
| 1 | $0.790 \pm 0.004$ | 10 | $0.629 \pm 0.011$ | $1.419 \pm 0.015$ |
|  |  | 20 | $0.873 \pm 0.018$ | $1.662 \pm 0.021$ |
|  |  | 30 | $0.950 \pm 0.017$ | $1.740 \pm 0.021$ |
| 1.5 | $0.793 \pm 0.004$ | 10 | $0.629 \pm 0.011$ | $1.422 \pm 0.015$ |
|  |  | 20 | $0.873 \pm 0.018$ | $1.665 \pm 0.021$ |
|  |  | 30 | $0.950 \pm 0.017$ | $1.743 \pm 0.021$ |
| 2 | $0.792 \pm 0.004$ | 10 | $0.629 \pm 0.011$ | $1.421 \pm 0.015$ |
|  |  | 20 | $0.873 \pm 0.018$ | $1.664 \pm 0.021$ |
|  |  | 30 | $0.950 \pm 0.017$ | $1.742 \pm 0.021$ |

Table 1